\documentclass{article} 
\def\d{\mbox{d}} 
\def\O{\Omega} 

\def\bra{\langle} 
\def\ket{\rangle} 
\def\a{\alpha} 
\def\b{\beta}
\def\dt{\delta}         
          
\def\L{\triangle}
          
\def\G{\Gamma} 
\def\e{\epsilon}

\def\lb{\lambda}

\def\m{\mu}
\def\n{\nu} 
\def\s{\sigma}

\def\p{\partial} 
\def\f{\frac} 
 
\def\l{\left}
\def\r{\right} 
\def\Pl{\ell_{\sm{Pl}}}
\renewcommand{\t}[1]{\tilde{#1}}
 
\newcommand{\sm}[1]{\mbox{\scriptsize #1}} 
\newcommand{\tn}[1]{\mbox{\tiny #1}}
\renewcommand{\@}[1]{\sqrt{#1}} 
\def\be{\begin{eqnarray}}
\renewcommand{\le}[1]{\label{#1}
\end{eqnarray}} 
\def\ee{\end{eqnarray}}
\newcommand{\eq}[1]{(\ref{#1})} 
\def\nn{\nonumber\\} 

\newcommand{\rf}[1]{\cite{ref:#1}}
\newcommand{\rr}[1]{\bibitem{ref:#1}} 
 
\def\smqu{\ {\buildrel ?\over =}\ }
\def\dts{\dt(\t\s-\t\s')} 
\def\ts{(\t\s-\t\s')}

\begin{document} 
\rm 
\large 
\rightline{THU-97/19}
\rightline{gr-qc/9707042} 
\LARGE\vskip 1.8in\centerline{Non-Commutative
Black Hole Algebra and String Theory from Gravity} 
\vskip 0.5in
\centerline{{\large S. de Haro}{\large \footnote{\large
e-mail: {\tt haro@fys.ruu.nl}}}} \vskip .3in\large 
\centerline{\it
Institute for Theoretical Physics} 
\centerline{\it University of
Utrecht} 
\centerline{\it Princetonplein 5, 3584 CC Utrecht}
\centerline{\it The Netherlands}\vskip .5cm

\centerline{\bf Class. Quantum Grav. 15 (1998) 519-535}


\section*{}

We generalize the action found by 't Hooft, which describes the
gravitational interaction between ingoing and outgoing particles in the
neighbourhood of a black hole. The effect of this back-reaction is that
of a shock wave, and it provides a mechanism for recovering information about
the momentum of the incoming particles. The new action also
describes particles with transverse momenta and takes into account the
transverse curvature of the hole, and has the form of a string theory action. 
Apart from the Polyakov term found by 't Hooft, we also find an antisymmetric 
tensor, which is here related to the momentum of the particles. At the quantum 
level, the identification between position and momentum operators leads to four
non-commuting coordinates.  A certain relation to M(atrix) theory is proposed.
\section*{}PACS numbers: 1125, 0460, 0470, 0470D, 0760

\section*{Introduction}

The $S$-matrix approach to the quantum black hole is an attempt to
solve the information paradox \rf{Hawk76}. This arose from Hawking's
discovery of black hole radiation and its thermal character. The claim
made by several researchers is that the inclusion of the gravitational
interaction between the ingoing particles and the Hawking radiation
that comes out of the hole will restore predictability; more exactly,
that in this way we will obtain a unitary mapping between an initial
pure state before the hole was formed and a pure final state after its
evaporation. Independently of whether this claim is true or not, it is
a fact that Hawking neglected the effect of these particles on the
metric, and that is in any case an important calculation to make, if
one wants to make accurate statements about black hole microscopy. For
one thing, these effects are not negligible if the particles have
Planckian energies, as they do when they fall into the hole.

So the first step was to compute the back-reaction of highly energetic particles 
near the horizon on the metric of the hole. Dray and 't Hooft found \rf{gnp85} 
that the effect is that of a shift in the position of the horizon, generally 
known as shockwave, and through this shift one had a mechanism to recover the 
information about the momentum state of the particles, essentially because the 
strength of the shift depends on the momentum. 't Hooft was then able to find 
the $S$-matrix describing the process \rf{gnp}. This turned out to be related to 
string theory, including the Nambu-Goto action of a Euclidean string (or, more 
properly, of a membrane at one instant in time) that is exchanged between the 
particles. These ideas were further studied in reference \rf{magg}, and made 
contact with the intuitive ``membrane paradigm'' in \rf{membrane}.

The next step after finding the $S$-matrix was to see what kind of
Hilbert space it worked on. This turned out to be a difficult problem,
with several technical complications. One of them was how to obtain a
discrete spectrum for this Hilbert space, with the right number of
microstates, according to the Bekenstein-Hawking entropy formula.
Another problem was the inclusion of the transverse gravitational
forces felt by the ingoing and the outgoing particles and their transverse
momenta, as all calculations had been done in the Rindler approximation. Some 
attempts at this were made \rf{g9402}, but not completely successful.

In this paper we will show that the covariant generalization of the
action appearing in the $S$-matrix is not the one previously advocated \rf{gnp}, 
but has to be slightly modified. The action one obtains is that of  bosonic 
string theory, including the antisymmetric tensor $B_{\m\n}$ with a field 
strength $H=\d B$, which turns out to be the Hodge dual of the momentum 
distribution. We will show that 't Hooft's equations for the shockwave can be 
seen as field equations arising from this string action. Our treatment is fully 
covariant, so that the particles are allowed to have momenta in any direction. 
When quantizing the theory, the equations of motion give rise to a set of four
non-commuting coordinates which nevertheless still depend on the string
degrees of freedom. We will argue that this problem could be cured by 
considering our description to be an effective one, so that the string is 
actually made up of particles coming in at definite positions, with a limited 
number, $N$, of them.  At the end a possible connection with M-theory is made.

The requirement of conformal invariance for this string theory will give us an 
expression for Einstein's equation which presumably contains solutions that 
describe a black hole interacting with matter. The detailed study of this is 
left for a future paper \rf{s}.

In appendix A this formalism is extended to an arbitrary number of dimensions.

\section{A covariant action for the gravitational interaction}

One of our aims is to include the transverse gravitational forces and transverse 
momenta in the $S$-matrix approach of 't Hooft. One would also like to improve 
the algebra obtained in \rf{g9402}, which gives an uncertainty relation between 
the light-cone coordinates when there are particles travelling at very high 
energies. Therefore we need the covariant generalization of the black hole 
action 
\be
S=\int\d\s\d\tau\l(T\,U\L V-P_{\sm{U}}U+P_{\sm{V}}V\r), 
\le{9}
with the definition $\L =\p_\s^2+\p_{\tau}^2$. $\s$ and $\tau$ are coordinates 
that parametrize the surface of the horizon. Since this action was derived in 
the Rindler approximation, they are set equal to the transverse coordinates $X$ 
and  $Y$ (in Schwarzchild spacetime, they correspond to the angles $\theta$ and 
$\phi$). $U$ and $V$ are light-cone coordinates, describing the location of the 
incoming and outgoing particles. The "string tension" $T$ is defined by 
$T=\f{1}{8\pi G}$, where $G$ is Newton's constant. Expression \eq{9} was found 
in reference \rf{gnato} to be the action appearing in the $S$-matrix for 
highly energetic scattering near a black hole. The $S$-matrix gives the momentum 
state of the outgoing particles once that of the ingoing particles is known, and 
{\it vice versa}. The covariant generalization must account for the relative 
minus sign between the momenta (the importance of this is shown in appendix B ), 
which comes from complex conjugation in the bra of the $S$-matrix. We expect 
these relative sign differences to appear when we include more general states
like $|P_{\sm{U}},P_{\sm{V}},P_{\sm{X}},P_{\sm{Y}}\ket$, and they will not be 
removable by a redefinition of coordinates. A general $S$-matrix element will be 
of the form 
\be 
_{\sm{out}} \bra
P_{\sm{U}}',P_{\sm{V}}',P_{\sm{X}}',P_{\sm{Y}}'|P_{\sm{U}},P_{\sm{V}},
P_{\sm{X}},P_{\sm{Y}}\ket _{\sm{in}}.  
\le{10} 
However, it is difficult to calculate this matrix element in the way it was done 
when one had the longitudinal directions only. Instead, we generalize 
covariantly the expression found by 't Hooft, and in particular the action.  
Thisis a much easier task to do.

All the ingredients we need for this are present in \rf{gnato}, but we will 
repeat the line of reasoning below. The arguments to be presented next were used 
in \rf{gnato} to improve the commutator between the longitudinal coordinates, 
from which one could then find an algebra for the surface elements 
\be W^{\m\n}=\e^{ij}\p_iX^\m\p_jX^\n.  
\le{10d}
These were combined in a certain way and integrated over some region of the 
horizon to give the SU(2) algebra. This had the advantage that one got rid of 
the $\t\s$-dependence, obtaining a discrete spectrum, see \rf{g9607}. In this 
paper, however, we will regard the covariant approach of \rf{gnato} as more 
fundamental, not only because it was used in the derivation of the algebra of 
the operators \eq{10d}, but also because it makes full contact with string
theory. As the main tools needed to covariantly generalize the action \eq{9} are 
the same one uses to derive the algebra of the surface elements \eq{10d}, the 
validity of our calculation is the same as that of the algebra.

So the first thing to take into account is the relative minus sign between the 
terms $P_{\sm{U}}U$ and $P_{\sm{V}}V$ in the action \eq{9}. For that purpose we 
dispose of the four-dimensional tensor $\e_{\m\n\sm{XY}}$. We use the convention 
$U=1$, $V=2$, $X=3$, $Y=4$ and $\e^{1234}=\e^{12}_{34}=-\e_{1234}=1$. 
$\e_{\m\n\sm{XY}}$ would automatically project $\m,\n$ to the values $1,2$, if 
contracted in the product 
\be 
\e_{\m\n 34}\,P^\m X^\n, 
\le{11}
giving us the result \eq{9}. But now because of covariance we have to replace 
the term $\e_{\m\n 34}$ by the full tensor $\e_{\m\n\a\b}$, so the two lower
indices have to be projected onto the indices corresponding to the membrane:  
\be 
\e_{\m\n 34}=\f{1}{2}\,\e_{\m\n\a\b}\,\e^{ij}\dt^\a_i\dt^\b_j, 
\le{12}
where $i,j$ run over the values $3,4$. In addition, we want the action to be
invariant under reparametrizations of the membrane coordinates, in order that 
$i,j$ can also be identified with the $1,2$-directions of spacetime. This will 
allow us to embed our string in spacetime as we wish. As 't Hooft remarked, this 
can be done be means of $\p_iX^\m=\dt^\m_i$, which holds in the Rindler 
approximation. So if this is the correct generalization, we get a factor 
\be
\f{1}{2}\,\e_{\m\n\a\b}\,\e^{ij}\p_iX^\a\p_jX^\b, 
\le{13}
which in the approximation reduces to $\e_{\tn{UV}\s\tau}$\footnote{$\s,\tau$ 
here stand for the transverse coordinates $\t\s$, not to be confused with 
indices running from 1 to 4.}, as seen. So inserting this back in \eq{11}, the 
total action becomes 
\be
S=-\f{1}{2}\int\d^2\t\s\,\l(T\,\@{h}h^{ij}\,g_{\m\n}\p_iX^\m\p_jX^\n
+\e_{\m\n\a\b}P^\m X^\n\,\e^{ij}\p_iX^\a\p_jX^\b\r).  
\le{15}

Notice that the only arguments we used here were covariance and that we must get 
the appropriate limit \eq{9} when we go back to the Rindler approximation.

Now, the variation $X^\m\rightarrow X^\m+\dt X^\m$ of \eq{15} yields the 
equation of motion 
\be 
\L X^\m=-\f{1}{T\@{h}}\,\e^{\m\n}_{\a\b}\,\e^{ij}\l(\f{3}{2}\,\p_iX^\a\p_jX^\b 
P_{\n} + X^\b\p_iX^\a\p_jP_{\n}\r), 
\le{16} 
where we have defined a Laplacian $\L=\f{1}{\@{h}}\,\p_i\@{h}\,h^{ij}\p_j$.  
Because our fields are distributions that should be integrated over some region 
of the horizon, we can use partial integration. The equation of motion becomes
\be 
\L X^\m=-\f{1}{2T\@{h}}\,\e^{\m\n}_{\a\b}\,\e^{ij}\p_iX^\a\p_jX^\b
P_{\n}.  
\le{16b} 
In the Rindler gauge, this equation clearly reduces to equation \eq{9a} in 
appendix B, which was found by 't Hooft. For simplicity, we have taken the 
metric $g_{\m\n}(X)$ to be constant, so our spacetime is Ricci-flat. In a more 
precise calculation, it of course has to be varied as well, and \eq{16b} 
receives a correction (see \eq{eqsmotall} in appendix A), but this is a good 
starting point for the cases we are interested in at the moment (where the 
metric is flat).

The physical meaning of the factor $\e^{ij}\p_iX^\a\p_jX^\b$ in the second term 
of \eq{15} is still unclear. We would like to be able to interpret it in terms 
of Fourier transforms of the bras and kets in the $S$-matrix, but that is not 
straightforward to do. What we seem to learn from this is that when there are 
particles travelling in all directions, the solutions of the Schr\"{o}dinger 
equation are not simply plane waves, but very complicated functions which depend 
on the geometry induced by these particles on the horizon.

As said, it is difficult to derive our action from first principles, as was done 
in the longitudinal case. But there is one important physical effect which we do 
see from \eq{16b} and that one also expected: the shift in an arbitrary 
direction $X^\m$ receives contributions from the particles travelling in {\it 
all} perpendicular directions, and not only from one of them. So now the ingoing 
particles will also affect each other, and the same for the outgoing ones. For 
example, two ingoing particles with momenta $P_{\sm{V}}$ and $P_{\sm{V}}'$, 
respectively, in the $V$-direction, and momenta $P_{\sm{X}}$ and $P_{\sm{Y}}'$, 
respectively, in the transverse directions, will interact with each other. This 
fact could not be found from an action like the one in appendix B (see equation 
\eq{9b}), since the summation over all perpendicular directions in \eq{16b} is 
essential.  So the distinction between ingoing and outgoing particles, typical 
of the Rindler gauge, has lost its fundamental meaning. It will manifest itself 
only in the sign of the momentum in the longitudinal directions, after we have
chosen a gauge. What is now important is to know the total momentum distribution 
in a certain direction.

Next we try to quantize the model. $P_\m$ was defined as the operator working on 
the Hilbert space that is canonically conjugated to $X^\m$, satysfying the 
following relation:  
\be
[P_\m(\t\s),X^\n(\t\s')]=-i\,\dt_\m^\n\,\dts.  
\le{5} 
This corresponds to the momentum of the particles. Commuting both sides of 
\eq{16b} with $X^\n$ and applying \eq{5}, we easily get
\be
{}[X^\m(\t\s),X^\n(\t\s')]=-i\,F^{\m\n}\ts, 
\le{18} 
where 
\be
F^{\m\n}\ts\equiv\f{\Pl^2}{2}\,\e^{\m\n}_{\a\b}\,\e^{ij}\p_iX^\a
\p_jX^\b f\ts, 
\le{19} 
$\Pl^2$ being Planck's length. $f$ is the Green function defined by 
\be
\L f(\t\s-\t\s')=-\f{1}{8\pi\,\@{h}}\,\dts.  
\le{20} 
This is exactly the
commutator postulated in \rf{gnato}, which has been derived here by covariant 
generalization of the action \eq{9}. The only requirement was that it must give 
us the right equations of motion (see \eq{9a} in appendix B) in the Rindler 
limit, because those are the only expressions whose correctness is without 
doubt. We did neglect a higher-order correction to \eq{18} coming from the fact 
that $F^{\m\n}$ is itself also an operator, and for simplicity considered
Ricci-flat metrics only. Notice that in the flat-space limit, there is a well 
defined Fourier transformation that relates the two. In a curved space, 
however, equation \eq{5} will receive ${\cal O}(\Pl^2)$ corrections, which, 
however, affect \eq{18} only by a term of ${\cal O}(\Pl^4)$.

This commutator can be inverted to give a relation between the momenta. After 
some algebra, we easily find 
\be
[P_\m(\t\s),P_\n(\t\s')]=i\Pl^{-2}\,\e_{\m\n\a\b}\f{W^{\a\b}}{W^2}
\,f^{-1}\ts, 
\le{20a2} 
where $W^2\equiv W_{\a\b}W^{\a\b}$.

There are several subtleties about the commutator \eq{5}. One can wonder whether 
$P$ is really the momentum canonically conjugated to $X$ or not. The main 
problem is that it is not defined in the usual way in field theory, like 
\be 
P_\m(\cdot,t)=\f{\dt L}{\dt(\p_t X^\m)},
\le{canmom} 
where $t$ is some time variable and the dot denotes all the fields $P$ may 
depend on. Rather, it was assumed to be a known function of the world-sheet 
variables $\t\s$ (this function is explicitly given in equation \eq{20f} of 
section 2), which is allowed if we work in the momentum representation, as we 
are doing (see \eq{10}). It was then inserted in the action in such a way that 
it gives the right equations of motion when we vary $X$ (so we did not vary the 
action with respect to $P$ because $P$ is a known function; it is not integrated 
over in the $S$-matrix, see equation \eq{20e} in section 2). This problem is
closely related to that of time and of finding a Hamiltonian formulation for the 
problem, and at present we do not know how to solve it.  However, for other 
forces different from gravity are switched off, we do know that \eq{5} must be 
true because $P$ enters in the Einstein equations as being the momentum of the 
photons. In a momentum representation, the latter is known and hence this 
picture is consistent. Therefore, the momentum must be conjugated to the 
position of the particle. It is probably true that this should be realised as a
constraint in the theory\footnote{We thank Steven Carlip for drawing this to our 
attention.}, but have not yet succeed doing this\footnote{One possibility for 
going to the Hamiltonian formalism is to Wick rotate one of the two world-sheet 
coordinates, identifying it with the time of the string; it is, however, not 
clear whether this is consistent with the shockwave interpretation. The 
interaction would then be by exchange of a string that is frozen at the horizon, 
evolving in time along one of the angular coordinates. A less exotic possibility 
is to add an extra, third dimension that accounts for the time evolution of the 
membrane in time. We believe that finding a Hamiltonian formalism will solve the
problem about momentum.}.

Let us write \eq{16b} for the transverse fields explicitly. We get 
\be 
T\,\@{h}\,\L
U&=&-W^{XY}\,P_{\sm{V}}+W^{UY}\,P_{\sm{X}}-W^{UX}\,P_{\sm{Y}}\nn
T\,\@{h}\,\L
V&=&+W^{XY}\,P_{\sm{U}}-W^{VY}\,P_{\sm{X}}+W^{VX}\,P_{\sm{Y}},
\le{20a3} 
with $W$ defined as in \eq{10d}. We thus indeed explicitly see that momenta in 
all directions contribute to the shift. These equations of course give us 
equation \eq{9a} in the limit.

The exact commutator is now 
\be
[U(\t\s),V(\t\s)]=-\f{i\Pl^2}{\@{h}}\,\e^{ij}\p_iX\p_jY\,f\ts,
\le{20b} 
which in the gauge $\t\s=(X,Y)$ becomes 't Hooft's commutator
\be 
[U(\t\s),V(\t\s)]=-i\Pl^2\,f\ts.  
\le{20b2}

It is important to see that our equations are manifestly covariant. For the 
transverse fields, we get from \eq{18} 
\be
[X(\t\s),Y(\t\s)]=-\f{i\Pl^2}{\@{h}}\,\e^{ij}\p_iU\p_jV\,f\ts,
\le{20c} 
so that, if we apply a Lorentz transformation on the Rindler gauge and take the 
membrane coordinates to be $\t\s=(U,V)$, we get 
\be
[X(\t\s),Y(\t\s)]=-i\Pl^2\,f\ts.  
\le{20d}

\section{The physical degrees of freedom}

One possible source of criticism is that in the Rindler gauge $\t\s=(X,Y)$, the 
commutator \eq{20c} is not well defined because on the right-hand side we have 
functionals depending on $\t\s$ and $\t\s'$, while the left-hand side would just 
be $[\s,\tau]$. But in this gauge, this commutator is not valid. From \eq{16b} 
we see that, because $\L X \sim P_{\sm{Y}}$ and $\L Y\sim P_{\sm{X}}$, if
$(X,Y)=(\s,\tau)$ the momentum in the transverse directions is zero,
$P_{\sm{X}}=0$ and $P_{\sm{Y}}=0$. So in that case the transverse coordinates 
are not physical fields, and their commutator vanishes. This is consistent with 
't Hooft's calculation, where the identification of the world-sheet and two of 
the target space coordinates forced the degrees of freedom to be reduced from 
four to two. Here we have decoupled these coordinates and allowed $X$ and $Y$ to 
be any function of $\s,\tau$, so we have a non-linear sigma-model with four
bosons living on a two-dimensional space. Therefore, up to a normalization 
constant, the matrix element \eq{10} is 
\be 
\bra
P_{\sm{out}}|P_{\sm{in}}\ket={\cal N}\int{\cal D}U(\t\s)\,{\cal
D}V(\t\s)\,{\cal D}X(\t\s)\,{\cal D}Y(\t\s)\,{\cal
D}h^{ij}(\t\s)\,\exp\,iS\,[U,V,X,Y,h^{ij}].  
\le{20e} 
Which functions are allowed for $X(\t\s)$ and $Y(\t\s)$ will follow from the 
equations of motion, just like for $U$ and $V$. For example, in the Rindler 
gauge, one usually takes 
\be
P_{\tn{U}}(\t\s)=\sum_{i=1}^Np_{\tn{U}}^i\,\dt(\t\s-\t\s^i)\nn
P_{\tn{V}}(\t\s)=\sum_{i=1}^Np_{\tn{V}}^i\,\dt(\t\s-\t\s^i), 
\le{20f2}
where $\t\s^i$ is the location of the $i$th particle. In this picture, $P(\t\s)$ 
is the total momentum distribution, which tells us how many particles are going 
in and out. It has a finite number, $N$, of contributions. $p^i$ is the momentum 
of each particle. Now we can do the same for the $X$ and $Y$ directions, taking 
generally 
\be
P_\m(\t\s)=\sum_{i=1}^Np_\m^i\,\dt(\t\s-\t\s^i).  
\le{20f} 
The solution that then follows from \eq{16b} is 
\be
X^\m(\t\s)=X_0^\m(\t\s)+\f{\Pl^2}{2}\sum_{i=1}^Np_\n^i\,
\e^{\m\n}_{\a\b}\,W^{\a\b}(\t\s^i)\,f(\t\s-\t\s^i), 
\le{20g}
$X_0^\m(\t\s)$ being a solution of the equation 
\be 
\L X_0^\m(\t\s)=0.
\le{freestring} 
So it is just the solution of the equations of motion of the free string, and 
can be expanded in Fourier modes $\a_n^\m$, as usual in string 
theory\footnote{Because it is an Euclidean wave equation, the boundary 
conditions are different from the usual ones. This will be explained 
elsewhere.}.

Notice that in \eq{20e} both the in and the out states contain particles moving 
in the $U$ and the $V$ directions. Here a difference from the original 
$S$-matrix arises. The labels ``in'' or ``out'' no longer have to do with the 
direction in which the particles are travelling. Rather, they represent the 
initial and final states of the particles, whatever their spatial configuration 
may be. The hole is the interaction region, an intermediate state that is 
integrated over in the path integral. This probably means that its mass has a 
complex component, as has been advocated in \rf{gnato}\rf{magg}. That is, we
think, the essence of the $S$-matrix description, but one will nor have
full understanding of this question until one has a Hamiltonian
formulation, because then one will have a picture with a preferred time
variable.

Like for 't Hooft's Lagrangian, one still has to impose a condition on the 
physical solutions which follows from the equations of motion of the world-sheet 
metric. The latter has also to be varied, because in \eq{20e} we integrate over 
all possible metrics. So the two-dimensional stress-energy tensor must vanish, 
\be 
T_{ij}=-\f{2}{T\@{h}}\,\f{\dt
S}{\dt h^{ij}}=0.  
\le{stressenergy} 
The result is the usual constraint for the metric on the string to be the metric 
induced by spacetime:
\be 
h_{ij}=g_{\m\n}\,\p_iX^\m\p_jX^\n.  
\le{eqmoth} 
This amounts to the condition that the positive frequency components of the 
Virasoro generators annihilate physical states. This brings us back to the
restriction that there are only $d-2$ physical bosons instead of $d$ (the 
spacetime dimension), like in 't Hooft's case, but now we can do this in a fully 
covariant way, imposing this condition at the end. In general, however, due to 
global effects and the presence of particles, the gauge will be more involved 
than just the Rindler gauge.

Classically, the stress-energy tensor \eq{stressenergy} is automatically 
traceless, without imposing the equations of motion. This is because, at the 
classical level, the action \eq{15} has conformal symmetry. Quantum 
mechanically, however, the trace can be non-vanishing if there are anomalies, 
which nevertheless can be cancelled by imposing very special conditions on, for 
example, the dimensionality of the spacetime. Since ---as we will next show--- 
our model is just a string theory, the same arguments used in critical string 
theory to demand this symmetry at the quantum level can be applied to our case. 
Therefore, in this paper we will not go into the rather involved discussion of 
whether or not anomalies can occur in more general cases, but will only consider 
the case that this conformal symmetry is maintained quantum mechanically.

As said, \eq{15} has a common interpretation in string theory, for which the 
equations that follow from conformal invariance are known. The action can be 
written in the following way:  
\be
S=-\f{1}{2}\int\d^2\t\s\,\l(T\,\@{h}h^{ij}\,g_{\m\n}\p_iX^\m\p_jX^\n
+B_{\m\n}(X)\,\e^{ij}\p_iX^\m\p_jX^\n\r), 
\le{20g3} 
which is exactly the action of a bosonic string propagating on a manifold with a
graviton background $g_{\m\n}(X)$ and an antisymmetric tensor field 
$B_{\m\n}(X)$. The antisymmetric tensor is, in our case, 
\be
B_{\m\n}(X)=\e_{\m\n\a\b}\,P^\a X^\b.  
\le{20g4} 
Hence the shockwave modifies the background where it propagates by giving it a 
non-vanishing torsion. The equations of motion, which shift the position of the 
particles in the perpendicular directions, are similar to the influence of a 
magnetic field on an electron in a cyclotron.

\section{A conformally invariant horizon}

The analogy with string theory discovered in \rf{gnato} is now found to
be more accurate: we have the same field content as in string theory, except for
the dilaton field. We could not derive its existence, which is only possible if,
from the beginning, one fully takes into account the curvature of the horizon, 
since the dilaton couples to the two-dimensional Ricci scalar, which was zero in 
't Hooft's calculation. Also, the dilaton does not have a clear physical 
interpretation in the black hole context yet, but we know it must be there, 
since it is part of the spectrum of the string that is obtained by a Virasoro 
decomposition of the fields satisfying the wave equation \eq{16}. The dilaton 
term is needed to ensure conformal invariance and finiteness at the quantum 
level, and so in this section we study the case that this conformal anomaly is 
indeed cancelled, although it remains a subtle issue how this invariance is 
realized (see the discussion in appendix A). Therefore we introduce by hand the 
following (non-Weyl-invariant) dilaton term to the action, in the usual way in 
string theory:  
\be
S_{\phi}=-\f{1}{4\pi}\int\d^2\t\s\,\@{h}\,\phi(X)\,R^{(2)}, 
\le{dil}
where $R^{(2)}$ is the two-dimensional Ricci scalar. For Ricci-flat metrics 
like, for example, $h_{ij}=\lb\,\dt_{ij}$, with constant $\lb$, this term is 
zero and thus consistent with 't Hooft's result.

The system \eq{20g3} with the extra term \eq{dil} has been well studied
\rf{callan}. The mentioned scale invariance turns out to require the following 
$\b$-functions for each of the fields to vanish\footnote{We are grateful to 
Yolanda Lozano for clarifying the analogy with string theory to us and 
suggesting to calculate the $\b$-functions.}:  
\be
\b^{\phi}
&=&\f{d-26}{48\pi^2}+\f{G}{4\pi^2}\l[4\,(\nabla\phi)^2-4\,\nabla^2\phi
-R+\f{1}{12}H^2\r]+{\cal O}\,(G^2)\nn
\b_{\m\n}^{g}
&=&R_{\m\n}-\f{1}{4}\,H_\m^{\a\b}H_{\n\a\b}+2\,\nabla_\m\nabla_\n
\phi+{\cal O}\,(G)\nn 
\f{1}{4G}\,\b^B_{\m\n}&=&\nabla_\a
H_{\m\n}^\a-2\,(\nabla_\a\phi)\,H_{\m\n}^\a+{\cal O}\,(G), 
\le{20g5}
where Newton's constant now plays the role of the string constant $\a'$ and 
$R_{\m\n}$ is the four-dimensional Ricci tensor. In our notation, $H^2\equiv 
H_{\m\n\a}H^{\m\n\a}$. The antisymmetric tensor field strength, $H_{\m\n\a}$, is 
defined by $H=\d B$, and, in this case, equals 
\be
H_{\m\n\a}=-3\,\e_{\m\n\a\b}\,P^\b, 
\le{20g6} 
hence it is the dual of the momentum. By definition, it satisfies the Bianchi
identity.

We see from \eq{20g5} that, in the presence of a dilaton, we are restricted to 
$d=26$ if we want to cancel the conformal anomaly, which is the usual result in 
bosonic string theory.

Notice that both the $\b$-function for the dilaton and that for the momentum 
torsion are linear in the Newton coupling constant. This is because their 
coefficient in the action is smaller by a power of $G$ than that of the Polyakov 
term. For the dilaton we just assumed that it had the same power of $G$ as in 
string theory. But for the momentum torsion tensor, this is a consequence of the 
equations of motion for the shockwave, since the momentum comes in with the 
first power of $G$ on the right-hand side of Einstein's equation. Therefore, as 
remarked in \rf{callan}, the {\it classical} contributions of the dilaton and
the momentum torsion to the anomaly are of the same order as the one-loop {\it 
quantum} contribution of the $g_{\m\n}$ coupling.

Equations \eq{20g5} are the equations of motion of the following effective 
action:  
\be
S_{\sm{eff}}=\int\d^dx\@{g}e^{-\phi}\l[R-\f{1}{12}H^2+4\,(\nabla\phi)^2\r],
\le{20g7} 
which, as shown in \rf{callan}, can be obtained from the Chapline-Manton 
\rf{chapline} supergravity action after rescaling of the four-dimensional 
metric.

If we would require the action to be supersymmmetric, which is not hard
to do, standard results of string theory would require the number of dimensions 
to be $d=10$. In that case, because 't Hooft's calculation is independent of the 
number of transverse dimensions, and the generalization is straightforward, the 
shock wave would not be a membrane (with time left out), but a Euclidean 7-brane 
(see Appendix A). The latter is also needed if we want to apply the results to 
Schwarzschild spacetime, where (in Rindler gauge) the membrane is identified 
with the horizon of the black hole\footnote{We thank Gijsbert Zwart for pointing 
this out to us.}. So to be consistent we have to regard our model as an 
effective description that arises after compacification down to 4 dimensions. 
Therefore it is important to have the full action, which after compactification 
reduces to \eq{20g3}. In fact, this action is easy to find:  
\be
S&=&-\;\;\,\f{T_d}{2}\int\d^{d-2}\s\l(\@{h}\,h^{ij}g_{\m\n}\,\p_iX^\m\p_jX^\n
-(d-4)\,\@{h}\r)
\nn
&&\nn
&-&\f{1}{(d-2)!}\int\d^{d-2}\s\,B_{\m_1\cdots\m_{d-2}}(X)\,\e^{i_1\cdots
i_{d-2}}\,\p_{i_1}X^{\m_1}\cdots\p_{i_{d-2}}X^{\m_{d-2}},
\le{dimension2} 
where the antisymmtric tensor is given in appendix A and the other results of 
section 1 are generalized to arbitrary dimension.

One of course would like to have also some physical motivation for these extra 
dimensions, but at present we are not in a position to say very much about this. 
It turns out \rf{g9607} that electromagnetic interactions can be included quite 
naturally in this formalism as a fifth Kaluza-Klein dimension, but it is hard to 
extend the theory to include the other interactions.

Combining the first and the second of equations \eq{20g5} and applying 
Einstein's equation 
\be 
R_{\m\n}-\f{1}{2}\,R\,g_{\m\n}=-8\pi
G\,T_{\m\n}, 
\le{20g10} 
gives us a stress-energy tensor equal to\footnote{One must not forget that there 
are higher-order corrections to these equations.} 
\be T_{\m\n}=\f{9}{2}\l(P_\m
P_\n-\f{1}{2}\,P^2g_{\m\n}\r)-2\,\nabla_\m\nabla_\n\phi
+2\,g_{\m\n}\nabla^2\phi-2\,g_{\m\n}\,(\nabla\phi)^2.
\le{20g11} 
The first part of this expression is reminding of the energy-momentum tensor of 
a perfect relativistic fluid, where $P_\m$ plays the role of the fluid velocity 
$u_\m$. This analogy will be studied in \rf{s}.

Finally, we seem to be back to a dynamical Einstein equation \eq{20g10}. This 
presumably describes black holes with matter falling in and going out and 
interacting gravitationally at the horizon \rf{s}. This is exactly what we 
wanted, because the aim of 't Hooft's calculation was to include the 
back-reaction on the metric. So the next step will be to look for physically 
sensible solutions of \eq{20g10} and then calculate Hawking's temperature (for a 
similar approach, in a somewhat different context, see \rf{gentropy}).

\section{A discrete algebra} 

The analogy with string theory is beautiful, but when considering the physical 
meaning of the theory one has to recall that equation \eq{20g3} only has a 
meaning as an effective action. In partucular, we would like to give the 
equations of motion and relation \eq{16b} a meaning in terms of single 
particles. The operators $X^\m(\t\s)$, as they are treated here, indeed give the
distribution of ingoing and outgoing particles on the surface of the horizon. 
Hence we have to go back to a discrete representation, where we have $N$ of 
these particles.

The first thing to remark (see \rf{gnp}) is that the physical Fock space of this 
theory is very different from ordinary Fock space, because in \eq{20f} the 
momentum distribution does not distinguish between different particles that are 
at the same position on the surface of the horizon. Thus, the total number of 
particles is not well defined in the usual sense or, more accurately, it is 
defined by the number of ``lattice sites''\footnote{We thank Leonard Susskind 
for the remark that this heuristic terminology should not be taken literally. 
These ``lattice points'' of course do not need to be fixed. They must rather be 
understood as the positions of the $0$-particles on the membrane 
\rf{gnp}\rf{matrix}.}. The discrepancy with usual field theory lies in the fact 
that in the low-energy limit, one is not interested in what happens when two 
particles are at exactly the same location, because the cross-sections of 
scattering processes, dominated by the low-energy interactions, are much larger 
than that. But when the gravitational force dominates, this question becomes 
relevant, even when the radial separations are not small.

So we have to apply \eq{20f} to the above results. To do this we will have to 
integrate with test functions that live on the horizon, in the following way:  
\be
\int\d^2\t\s\,F(\t\s)\,I(\t\s)=\sum_{i=1}^NF^i\,I(\t\s^i), 
\le{20h2}
where $F$ can carry any other spacetime index and $I$ is an arbitrary function.  
These distributions thus satisfy 
\be
F(\t\s)=\sum_{i=1}^NF^i\,\dt(\t\s-\t\s^i).  
\le{20h4}

On the other hand, for large $N$ we have 
\be
\int\d^2\t\s\,F(\t\s)\simeq\f{A}{N}\sum_{i=1}^NF(\t\s^i), 
\le{20m}
where $A$ is the area of the horizon, $A=16\pi M^2$. Thus we have 
\be
F^i=\f{A}{N}\,F(\t\s^i).  
\le{20h3}

{}From this and equation \eq{20a2} one can get an algebra for the momentum of 
each particle. To do this we again go to the Rindler approximation, where we 
identify two spacetime directions with $\s$ and $\tau$. So in that 
approximation, 
\be
\f{1}{2}\,\e_{\m\n\a\b}\,\e^{ij}\p_iX^\a\p_jX^\b=\e_{\m\n\s\tau}\equiv\e_{\m\n},
\le{20h} 
where the indices $\m$, $\n$ are now restricted to the longitudinal plane, and
$W_{\a\b}W^{\a\b}=2$.

Writing down equation \eq{20a2} for the specific case $\t\s=\t\s^i$, 
$\t\s'=\t\s^j$, we get
\be
[P_\m(\t\s^i),P_\n(\t\s^j)]=i\Pl^{-2}\,\e_{\m\n}\,f^{-1}(\t\s^i-\t\s^j);
\le{20j}
now comparing \eq{20f} with \eq{20h4}, and using equation \eq{20h3}, we get
\be
[p_\m^i,p_\n^j]=i\e_{\m\n}\l(\f{16\pi M^2}{N}\r)^2f^{-1}(\t\s^i-\t\s^j).  
\le{20j2} 
For example, for a large black hole ($M\rightarrow\infty$) in four dimensions, 
$f$ is approximately given by 
\be
-\f{1}{2T}\log\f{\|\t\s^i-\t\s^j\|^2}{\Pl^2}, 
\le{20j6} 
for any two particles $i$ and $j$. Now if we assume all particles to be 
homogeneously distributed throughout the horizon, the mean distance between two 
of them will be $\|\t\s^i-\t\s^j\|\simeq \@{A/N}=\@{16\pi M^2/N}$, and so 
\be 
[p_\m^i,p_\n^j]_{i\not= j}=-i\,\e_{\m\n}\,\f{32\pi M^4}{GN^2\log\f{16\pi
M^2}{N\Pl^4}}.  
\le{20j7}

We of course consider this discrete model, especially the last part, where we 
did a kind of mean-field theory, as a toy model, purely as an indication of the 
direction in which one has to search for an algebra with a finite number of 
degrees of freedom. Nor did we derive that the number of particles is finite or 
that there is something like a minimal length on the horizon. However, in 
spacetime a minimal length scale {\it does} explicitly appear in equation 
\eq{18}. The unusual fact about this is that it depends on the location of the 
particles on the world-sheet by means of the propagator $f\ts$.

\section{M(atrix) theory and gravity} 

At first sight, the commutator \eq{18} looks very much like what is found in 
M(atrix) theory, because it is proportional to the orientation tensor \eq{10d} 
(the ``Poisson bracket'' of the membrane): it is its Hodge dual. However, one 
must be very careful. This commutator comes from Dirac quantization, and not
from a matrix representation of the fields, like in matrix theory. The striking 
fact is that the coordinates do not commute, not in the sense of matrix theory, 
where the matrices are non-commuting but the matrix elements themselves are 
numbers, but in the sense of quantum mechanics; our prescription is really a 
Dirac quantization condition. The commutator \eq{18} gives us a rule for making 
the transition from the classical to the quantum Euclidean string (or membrane 
at one instant of time) in the presence of highly energetic particles. So, 
though at a speculative stage, and although the noncommutativity has a different 
origin here and in matrix theory, we wish to push the analogy forward to see if 
we can learn something about matrix theory from this model.

In matrix theory, one has the representation\footnote{We use the following 
notation: $x^\a$ are the classical functions representing $X^\a$, which can be 
expanded in a truncated summation over membrane modes; the M in 
$[\cdot]_{\sm{M}}$ stands for ``matrix commutator''.}
\be
\f{1}{N}\,\{\cdot\}\leftrightarrow [\cdot]_{\sm{M}},
\le{21a}
where $N$ is the cutoff imposed to regularize the membrane. Now in our case,
the quantum mechanical commutator is given by calculating the dual of the 
Poisson bracket, 
\be
i\Pl^2\,\e^{\m\n}_{\a\b}\,\{x^\a,x^\b\}\,f,
\le{21}
{\it i.e.}, quantization is determined by the replacement 
\be
\ast\,\{\cdot\}\rightarrow \f{1}{i\Pl^2 f}\,[\cdot],
\le{22} 
$\ast$ denoting the Hodge dual\footnote{We now work in 4 dimensions, see 
appendix A for a generalization.}. So $\Pl^2f$ seems to play the role of 
$\f{1}{N}$ in matrix theory. It also gives a minimal length scale in spacetime. 
One is therefore tempted to make the identification\footnote{Notice that the
truncation of the series then depends on $\|\t\s-\t\s'\|$. We are not sure that 
this is a reasonable assumption, but, as already remarked, the cutoff of 
spacetime turns out to depend on the propagator on the world-sheet. A more 
conservative viewpoint is to take simply $N=\f{1}{\ell_{\mbox{\tiny Pl}}^2}$.} 
$\Pl^2f\equiv\f{1}{N}$, or $G\sim\f{1}{N}$; the Fourier expansion would then be 
truncated like $N\sim \f{1}{\ell^2_{\sm{Pl}}}$. Physically, this means that the 
membrane is naturally regularized by Planck's length if one takes the momenta of 
the particles into account. So we propose the following series:
\be
\f{1}{N}\ast\{\cdot\}\;\leftrightarrow\;\ast\,[\cdot]_{\sm{M}}\;\rightarrow
\;\f{1}{i\hbar}\,[\cdot].  
\le{24} 
The first correspondence is the matrix representation of membrane theory, with 
the specific choice $N\sim\f{1}{G}$; the second is Dirac's quantization 
condition. If our proposal is correct, roughly speaking we have one degree of 
freedom per Planck area, as 't Hooft has suggested in his brick-wall model. At 
large distances, one of course recovers the commutative limit 
$N\rightarrow\infty$, meaning that the particles form a continuum and the 
horizon behaves classically.

Also the Fock space of our theory resembles very much that of matrix theory. We 
have interpreted the membrane as made up of a finite number, $N$, of particles.  
So the membrane description is only an effective one, which appears in the 
large-distance limit. These ideas have long been advocated by 't Hooft in, for 
example, \rf{gnp}.

It is clear that very much has to be done to understand and check, if possible, 
this analogy, in particular the choice $N\sim\f{1}{\Pl^2}$, as well as to 
understand the M(atrix) theory proposal itself. In particular, although the 
factor $N$ explicitly appears in our formulae, suggesting that in the limit 
$N\rightarrow\infty$ both the left- and right-hand sides of \eq{21a} are zero, 
in principle one can get rid of this $N$-dependence by a redefinition of the 
fields $X$. But in that case, the generators of the group diverge and the 
membrane becomes an infinite plane. It is unclear whether this picture is 
physically meaningful\footnote{This is analogous to rescaling the operators $x$ 
and $p$ of quantum mechanics, that satisfy $[x,p]=i\hbar$, by a factor of 
$\@{\hbar}$: $X\equiv \f{x}{\@{\hbar}}$, $P\equiv\f{p}{\@{\hbar}}$, so that they 
obey $[X,P]=i$ even in the limit $\hbar\rightarrow0$; in that limit, however, 
they diverge, and the physical interpretation as the position and momentum of a 
particle is in any case different; the noncommutativity in the classical limit 
is rather an artifact of blowing up a very small region of phase space by an 
infinite factor. This is similar to what happens in the case of the membrane. 
Therefore, one has to be careful about these rescalings.}, and the dependence on 
the choice of basis of the M(atrix) theory proposal has to be studied as well. 
Also, the commutator \eq{18} receives a higher-order correction which could 
modify the resemblance to matrix theory. Note \rf{bernh}, however, that, in the 
same way, for large but finite $N$ there are higher order corrections to the 
relation between the algebra of area-preserving diffeomorphisms and the SU(N) 
algebra, so that only in the limit are both goups the same.

\section*{Conclusion}

One of the questions one would ultimately like to answer\footnote{We are 
grateful to Finn Larsen for this remark.} concerns the number of microstates of 
the hole. Although a trial to get some insight into this has been made in 
section 4, we are still far from an answer. In particular, one would like to be 
able to calculate the entropy. Yet something can be learned directly from the 
action appearing in the $S$-matrix. The first term is the Polyakov term, which, 
imposing the constraint $h_{ij}=g_{\m\n}\,\p_iX^\m\p_jX^\n$, becomes the 
Nambu-Goto term, namely, the horizon area. More generally, it describes the 
world volume of the $p$-brane. So this suggests that, in any dimension, 't 
Hooft's action, which is Euclidean\footnote{However, there is a factor of $i$ in 
front of the action which we do not understand.}, is related to the entropy as 
far as the first term is concerned. It is possible that the second term gives 
corrections to the entropy due to the presence of ingoing and outgoing particles 
(in fact, the equation of motion \eq{16b} describes the oscillations of the
membrane, whose area increases with the ingoing particles and decreases
with the outgoing ones), and that these corrections can be included in the 
area-term like in a Born-Infeld action, but more work is needed in this 
direction. Another possible approach is to calculate the entropy from the 
Einstein equations one obtains in the low-energy limit.

There are other subtle points about 't Hooft's $S$-matrix, especially when one 
considers realistic, "astrophysical", black holes \rf{itzhaki}. One, however, 
has to realize that the $S$-matrix obtained so far cannot be the whole story, 
but is a crude approximation to a microscopic description of the horizon. 
Nevertheless we did learn something from this microscopic model. The dynamics 
of particles at high energies, which carry shockwaves with them, can be 
described by a string theory action in a background including a torsion, which 
is related to the momentum of the particles. Furthermore, the coordinates of 
these particles obey the uncertainty principle, with an uncertainty proportional 
to Newton's constant.

\section*{Acknowledgements}

I would like to thank G 't Hooft for his useful suggestions and discussions and 
Y Lozano for suggesting calculating the $\b$-functions and clarifying the 
relation with string theory. I also thank S Carlip for his useful remarks about 
quantization, and M Maggiore, F Larsen, E Verlinde, G Arcioni and V  Pascalutsa 
for interesting discussions and suggestions. C O Lousto provided me with two of 
the references. This work has been supported by a Spinoza grant.

\appendix

\section*{Appendix A. Generalization to $d$ dimensions}

In this appendix we generalize some results of sections 1 and 2 to any dimension 
$d>4$. This will be useful if one wants to work in the critical dimensions 
$d=26$ or $d=10$. First of all, the antisymmetric tensor appearing in 
\eq{dimension2} is equal to 
\be
B_{\m_1\cdots\m_{d-2}}(X)=\e_{\m_1\cdots\m_{d-2}\a\b}\,P^\a X^\b, 
\le{dimB} 
so the action is 
\be
S&=&-\;\;\,\f{T_d}{2}\int\d^{d-2}\s\l(\@{h}\,h^{ij}g_{\m\n}\,\p_iX^\m\p_jX^\n
-(d-4)\,\@{h}\r)\nn&&\nn
&-&\f{1}{(d-2)!}\int\d^{d-2}\s\,\e_{\m_1\cdots\m_{d-2}\a\b}\,P^\a
X^\b\,\e^{i_1\cdots
i_{d-2}}\p_{i_1}X^{\m_1}\cdots\,\p_{i_{d-2}}X^{\m_{d-2}}.
\le{dimension}

The field-strength $H=\d B$ is then given by 
\be
H_{\m_1\cdots\m_{d-1}}=-(-1)^d\,(d-1)\,\e_{\m_1\cdots\m_{d-1}\a}\,P^\a.
\le{dimH} 
The equations of motion of \eq{dimension} are \rf{duff} 
\be
\p_i\l(\@{h}\,h^{ij}\,g_{\m\n}\,\p_jX^\n\r)+h^{ij}\,g_{\m\n}\,
\G^\n_{\;\;\a\b}
\,\p_iX^\a\p_jX^\b=\f{1}{(d-1)!\,T_d}\,H_{\m\m_1\cdots\m_{d-2}}\,
W^{\m_1\cdots\m_{d-2}}, 
\le{eqsmotall} 
where the orientation tensor is now 
\be 
W^{\m_1\cdots \m_{d-2}}=\e^{i_1\cdots
i_{d-2}}\,\p_{i_1}X^{\m_1}\cdots\p_{i_{d-2}}X^{\m_{d-2}}.
\le{dimorten} 
For a flat spacetime metric, this reduces to 
\be 
\L X^\m=\f{1}{(d-1)!\,T_d\,\@{h}}\,H^\m_{\;\,\m_1\cdots\m_{d-2}}\,W^{\m_1\cdots
\m_{d-2}}, 
\le{dimeqsmot} 
which agrees with the supergravity result obtained in the appendix of 
\rf{becker} for the shock wave in $d=11$. Requirement \eq{5} gives us 
\be
[H_{\m_1\cdots\m_{d-1}}(\t\s),X^\n(\t\s')]=\f{i(-1)^d\,(d-1)}{\@{h}}\,
\e_{\m_1\cdots\m_{d-2}}^{\;\;\;\;\;\;\;\;\;\;\;\;\;\;\n}\,\dts
\le{commHX} 
and thus 
\be
[X^\m(\t\s),X^\n(\t\s')]=-i\Pl^2\,\f{1}{(d-2)!}\,
\e^{\m\n}_{\;\;\;\,\m_1\cdots\m_{d-2}}\,W^{\m_1\cdots \m_{d-2}}\,f\ts.
\le{dimcomm} 
Notice that for $d\not= 4$, the propagator is not a logarithmic function, but 
(for a flat world-sheet metric) 
\be
f\ts=-\f{1}{T_d\,(d-4)\,\O_{d-2}}\,\f{1}{\|\t\s-\t\s'\|^{d-4}},
\le{dimprop} 
where $\O_{d-2}$ is the surface area of a unit sphere in $d-1$ dimensions.

A possible problem is that, for our model, conformal invariance only occurs in 4 
dimensions, because only then is the horizon a Euclidean string. This is at 
first sight inconsistent with the fact that conformal invariance requires 
$d=26$, or $d=10$ after adding fermions. However, conformal invariance 
re-emerges when letting the compactification radius go to zero (see reference 
\rf{stelle} for conformal invariance arising from the usual diffeomorphism 
invariance of the membrane, after double-dimensional reduction). What happens in 
the presence of Kaluza-Klein modes or winding in the compactified dimensions, 
and whether conformal invariance arises after compactification as a 
manifestation of some other symmetry of the higher-dimensional theory, is still 
unknown. This will be subject of a future investigation.

\section*{Appendix B. Uniqueness of the covariant action}

In this appendix we argue that the expression which should reduce to equation 
\eq{9} in the appropriate limit and has until now been used as its covariant 
generalization should be slightly modified, in the way done in section 1.

The earlier proposed candidate \rf{gnp} is 
\be 
S\smqu\int\d\s\d\tau
\l(-\f{T}{2}\,\@{h}h^{ij}g_{\m\n}\,\p_iX^\m\p_jX^\n+P_\m(\s,\tau)\,
X^\m(\s,\tau) \r), 
\le{1} 
where $h_{ij}$ is the metric on the world-sheet, $h\equiv\det h_{ij}$, 
$g_{\m\n}$ is the four-dimensional metric in target space, $\t\s=(\s,\tau)$ are 
the coordinates on the world-sheet, and the string tension is $T=\f{1}{8\pi G}$. 
Notice that this action is generally covariant in the world-sheet and in target 
space coordinates, and under reparametrizations of $h_{ij}$. $P_\m$ now 
transforms as a density in the world-sheet coordinates (in contrast to our 
convention of section 1).

The limit without transverse momenta and curvature is then obtained taking a 
conformally flat metric on the world-sheet, 
\be
h_{ij}=\lb(\s,\tau)\,\dt_{ij}, 
\le{1b} 
for arbitrary $\lb$, and working in the Minkowski approximation to Kruskal 
coordinates, which holds at points near to the horizon, where $r\simeq 2M$:  
$X^\m=(U,V,X,Y)$. The metric is then 
\be 
\d s^2=2\,\d U\d V+\d X^2+\d Y^2.  
\le{2} 
Up to a sign redefinition of one of the coordinates, \eq{1} gives us the same
equations of motion as \eq{9}.

But we now show that the covariant action \eq{1} cannot be consistently 
quantized. Variation of the action gives\footnote{Notice that throughout this 
paper we take the metric tensor to be constant, which is a good approximation in 
the cases we are interested in at the moment.} 
\be 
P_\m\smqu -Tg_{\m\n}\L X^\n.  
\le{9b} 
Now we would like to promote the $p$'s, which come from Fourier transforming the
$X^\m$-fields in the $S$-matrix, to operators that satisfy \eq{5}. But if we 
impose this equation on \eq{9b}, we get 
\be
g_{\m\n}[X^\n(\t\s),X^\a(\t\s')]\smqu-i\Pl^2\,g_\m^\a f\ts
\le{9c} 
and thus, contracting with $g^{\m\b}$, 
\be
[X^\b(\t\s),X^\a(\t\s')]\smqu-i\Pl^2\,g^{\a\b}f\ts, 
\le{9d}
which obviously cannot be true, independently of what redefinitions we take or 
what coordinate system we choose, because the metric tensor is symmetric and the 
commutator antisymmetric with respect to the indices $\m,\n$. In particular, 
choosing light-cone coordinates and redefining momenta with a minus sign will 
not help; so the conclusion is that the above action is not consistent with 
covariant quantization. Remarkably enough, for 't Hooft's model (equation 
\eq{9}) {\it was} consistent with it. This can only mean that the action was not 
generalized in the right way.

We will now choose a particular gauge to show where the discrepancy with 't 
Hooft's model comes from. In the Rindler gauge, \eq{1b} and \eq{2}, the action 
becomes 
\be 
S\smqu\f{T}{2}\int\d^2\t\s\l(2U\L V+X\L X+Y\L
Y\r)
+\int\d^2\t\s\l(P_{\sm{U}}U+P_{\sm{V}}V+P_{\sm{X}}X+P_{\sm{Y}}Y\r),
\le{3} 
where $\L\equiv\p_\s^2+\p_{\tau}^2$. Varying with respect to the four fields 
$U,V,X$ and $Y$, we find the equations of motion 
\be
P_{\sm{U}}(\t\s)&\smqu&-T\,\L V(\t\s)\nn P_{\sm{V}}(\t\s)&\smqu&-T\,\L
U(\t\s), 
\le{4} 
and for the transverse fields 
\be
P_{\sm{X}}(\t\s)&\smqu&-\f{T}{2}\,\L X(\t\s)\nn
P_{\sm{Y}}(\t\s)&\smqu&-\f{T}{2}\,\L Y(\t\s).  
\le{4b} 
This looks very much like a covariant generalization of 't Hooft's shift 
equations because the particles are allowed to have a transverse momentum ---but
it is not! Notice that $P_{\sm{X}}$ is proportional to $X$ and $P_{\sm{Y}}$ 
proportional to $Y$. In the original derivation of \rf{gnp85}, the shifts on the 
outgoing particle were in the direction perpendicular to the direction of the 
ingoing particle (in the $U$-$V$ plane), as above. So if we give a particle a 
momentum in the $X$ direction, the other particle must be shifted in the $Y$
direction\footnote{Actually, the shift will be in all the perpendicular
directions, but we consider one single plane for simplicity. This has been 
pointed out in section 2.}, having a relation like $P_{\sm{X}}\sim\L Y$ and 
$P_{\sm{Y}}\sim\L X$, not as in equation \eq{4b}. Next we present an argument to 
see that our Lagrangian is indeed inconsistent with covariant quantization, and 
that the coupling between the directions $U$ and $V$ in the equations of motion 
\eq{4} is not an artifact of the light-cone gauge. Following 't Hooft 
\rf{g9607}, we promote the $p$'s, which come from Fourier transforming the 
$X^\m$-fields in the $S$-matrix, to operators that are canonically conjugated to 
$X^\m$. We are allowed to do so, because the way they came in the calculation 
was as a parameter that determines the momentum of the particle. Therefore they 
must be conjugate to the position operators and satisfy \eq{5}. But one directly 
sees that the relation \eq{5} cannot coexist with \eq{4} and \eq{4b}.

To understand this, let us adopt the covariant notation \rf{g9607}. Up to an 
irrelevant overall minus sign, the only possibility is to define:  
\be X^+=V\nn X^-=U 
\le{6} 
and 
\be
P^-&=&P_+=\;\;\,P_{\sm{V}}\nn P^+&=&P_-=-P_{\sm{U}}.  
\le{7} 
Like in field theory, we have taken outgoing momentum with a minus sign, so
that all momenta are ingoing. This was needed in order to get non-vanishing 
commutators. But then $[P_{\sm{U}},U]=-i$, $[P_{\sm{V}},V]=-i$, gives us 
\be
{}[P_+(\t\s),X^+(\t\s)]\smqu-i\,\dts\nn
{}\;[P_-(\t\s),X^-(\t\s)]\smqu+i\,\dts, 
\le{8} 
which is inconsistent with \eq{5}! Otherwise we are forced to set 
$P_{\sm{U}}=U=P_{\sm{V}}=V=0$. The same happens if we try to quantize
the operators in the $X,Y$-directions, the inconsistency being much
more clear because of the proportionality $P_{\sm{X}}\sim X$,
$P_{\sm{Y}}\sim Y$. We then simply get $X=P_{\sm{X}}=Y=P_{\sm{Y}}=0$,
so that no transverse momentum is allowed.

Up to overall minus signs, the definition \eq{7} is the only
possibility\footnote{We indeed see that if we redefine $U$ with a minus sign, we 
also have to modify the conjugate momentum $P_{\sm{U}}$ (and {\it vice 
versa}).}.  So the action \eq{9} {\it is} consistent with
quantization, but if we go over to the covariant notation, the generalization is 
not \eq{1}.

Now the correct equations of motion are 
\be 
P_{\sm{U}}&=&+T\,\L V\nn
P_{\sm{V}}&=&-T\,\L U, 
\le{9a} 
which are essentially different from \eq{4}. Covariant quantization is then 
reached defining 
\be X^+&=&V\nn
X^-&=&U\nn P_+&=&P_{\sm{V}}\nn P_-&=&P_{\sm{U}}, 
\le{9a2} 
so that we obtain 
\be 
[U(\t\s),V(\t\s')]=-i\Pl^2\,f\ts.  
\le{9a3} 
This is, indeed, what is obtained if one uses the definitions of section 1.

\end{document}